\documentclass[sigconf, screen]{acmart}

\AtBeginDocument{%
  }

\setcopyright{none} 
\copyrightyear{$^\ast$ Corresponding author}
\acmYear{}
\acmDOI{}

\acmConference[OASIS-AI]{FlagOS Forum at MICRO 2026 - Open Compute Architecture and Software Infrastructure for Heterogeneous AI Systems}{November 01, 2026}{Athens, Greece}

\acmISBN{}

\usepackage{subcaption}
\usepackage[all]{nowidow}

\begin{document}

\title{How Well Does Graph Superoptimization Transfer? Adapting Tensor-Graph Superoptimizer to a~Triton-based NPU}
\subtitle{\normalsize{FlagOS Forum at MICRO 2026 - OASIS-AI: Open Compute Architecture and Software Infrastructure for Heterogeneous AI Systems}}
\author{Weronika Hryniewska-Guzik$^\ast$}
\affiliation{%
  \institution{Heterogeneous Memory Software Lab\\
  Huawei Technologies, Poland}}
\email{weronika.hryniewska.guzik@h-partners.com}

\author{Mateusz Comblik}
\affiliation{%
  \institution{Heterogeneous Memory Software Lab\\
  Huawei Technologies, Poland}}
\email{mateusz.comblik@h-partners.com}

\author{Daniel Murawski}
\affiliation{%
  \institution{Heterogeneous Memory Software Lab\\
  Huawei Technologies, Poland}}
\email{daniel.murawski3@huawei.com}

\author{Jakub Schmiegel}
\affiliation{%
  \institution{Heterogeneous Memory Software Lab\\
  Huawei Technologies, Poland}}
\email{jakub.schmiegel@huawei.com}

\author{Maciej Świetlik}
\affiliation{%
  \institution{Heterogeneous Memory Software Lab\\
  Huawei Technologies, Poland}}
\email{maciej.swietlik@h-partners.com}

\author{Maciej Kokociński}
\affiliation{%
  \institution{Heterogeneous Memory Software Lab\\
  Huawei Technologies, Poland}}
\email{maciej.kokocinski1@huawei.com}

\author{Tadeusz Kobus}
\affiliation{%
  \institution{Heterogeneous Memory Software Lab\\
  Huawei Technologies, Poland}}
\email{tadeusz.kobus@huawei.com}




\begin{abstract}
For automatic optimization systems, performance portability across increasingly diverse AI accelerators remains a challenge. We investigate how well graph-level superoptimization transfers across architectures by adapting Mirage from its CUDA-oriented design to the Huawei Ascend 910B3 through a Triton-based compilation stack. We evaluate the adapted system on transformer-relevant workloads, including matrix multiplication, softmax, RMSNorm+MatMul, gated MLP, FNN, and attention. The results show that graph-level optimization can successfully transfer and outperform available reference implementations by up to 87.0\% on fused workloads. However, performance is highly dependent on tensor shape and workload structure, with substantial regressions for other configurations. We identify target-specific tensor partitioning, execution-resource modeling, backend lowering and fusion, and search and verification costs as key limitations. Our findings demonstrate that a portable graph representation and programming model do not imply a portable optimization space, motivating modular target-specific mechanisms within otherwise reusable compiler infrastructure.
\end{abstract}



\maketitle
\renewcommand{\shortauthors}{Weronika Hryniewska-Guzik et al.}

\section{Introduction}

The growing variety of domain-specific AI accelerators has made performance portability a significant challenge for machine learning systems. Modern AI workloads are deployed on heterogeneous processors, whose execution models, memory hierarchies, and parallel resources can vary substantially, even though they provide similar high-level tensor programming abstractions~\cite{park2024nest,lopoukhine2025multi}. Compiler frameworks such as Triton~\cite{triton} aim to reduce this gap by providing a higher-level programming model for expressing efficient tensor kernels while allowing hardware-specific compilation to be handled by dedicated backends~\cite{zhu2026triton}. However, a common programming abstraction does not necessarily imply that the optimization strategies developed for one accelerator remain effective on another~\cite{lopoukhine2025multi, feng2023tensorir}.

This distinction is particularly important for automatic optimization systems. Rather than relying solely on manually engineered kernels, automatic tensor-program optimizers search over alternative program transformations, schedules, and mappings to discover efficient implementations~\cite{zheng2020ansor,wu2025mirage}. Mirage~\cite{wu2025mirage} is a representative example: it introduces $\mu$Graphs, a hierarchical representation of tensor programs spanning kernel-, threadblock-, and thread-level computation, and combines graph transformations with schedule optimization and custom-kernel generation. Its search procedure is explicitly structured around the execution hierarchy of modern GPUs, and Mirage demonstrates substantial performance improvements over existing approaches even on widely used and heavily optimized neural-network workloads~\cite{wu2025mirage}. However, an NPU represents a fundamentally different level of hardware specialization than a GPU, with substantially different execution and memory architectures, making direct transfer of GPU-oriented optimization strategies particularly challenging.

\begin{figure*}
    \centering
    \includegraphics[width=\linewidth]{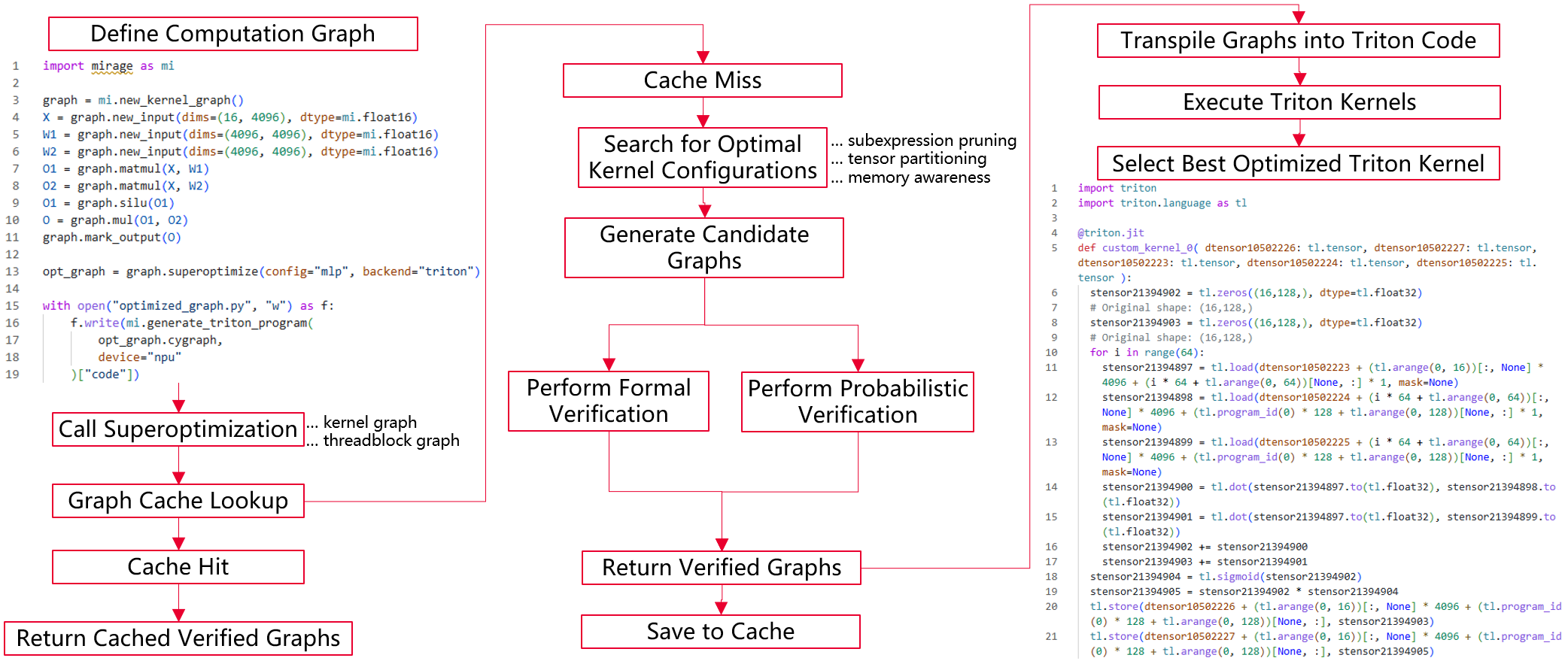}
    \caption{Architecture of the adapted graph superoptimization and Triton code generation pipeline.}
    \label{fig:pipeline}
\end{figure*}

The central research question of this work is therefore: \emph{how much of a graph superoptimizer's optimization capability transfers across accelerator architectures?} In this work, we distinguish functional portability from performance portability. Functional portability refers to the ability to compile and execute the generated programs correctly on a different device. Performance portability is stronger: the optimizer must also make effective decisions about partitioning, parallelism, memory placement, fusion, and scheduling for the target architecture.

We investigate this question by adapting Mirage to the Huawei Ascend 910B3 through the Triton programming stack. This setting allows us to distinguish between optimization mechanisms that transfer through the common graph representation and those that remain dependent on the original CUDA execution model. The adaptation requires changes to the compiler infrastructure, including operator coverage, verification, and integration with modern Triton versions.

We evaluate the resulting system on transformer-relevant workloads covering matrix multiplication, unscaled softmax, fused RMSNorm+MatMul, gated MLP, FNN, and attention. The results show that the adapted superoptimizer discovers competitive and sometimes substantially faster kernels, but its performance varies strongly with workload shape and structure. We identify target-specific tensor partitioning, execution-grid selection, backend-level lowering, and search cost as the main factors limiting consistent performance.

In summary, this paper makes the following contributions:
\begin{itemize}
\item A graph-level tensor superoptimiser has been adapted for the Huawei Ascend 910B3 processor via a code generation pipeline based on the Triton library, extending the operator, verification and compiler infrastructure required by the target stack.
\item Graph-level superoptimisation was evaluated for individual operators and multi-operator transformer workloads on a non-CUDA accelerator.
\item The gap between functional and performance portability was discussed, and specific limitations were identified regarding tensor partitioning, execution grid selection, backend-level fusion, and search and verification.
\item Design suggestions were proposed for separating reusable graph-level optimisation infrastructure from accelerator-specific optimisation mechanisms.
\end{itemize}

\section{Adapting graph superoptimization to a Triton-based NPU stack}

Figure~\ref{fig:pipeline} provides an overview of the adapted Mirage pipeline, from computation-graph construction and candidate generation through Triton code generation, verification, caching, and execution on the Ascend NPU.

\subsection{Removing CUDA-centric dependencies}

The original Mirage installation and build process contained dependencies on the CUDA software stack, regardless of the backend used. Therefore, these dependencies are currently separated. The Triton dependency has been modernized to support the Triton versions required by the Ascend backend (3.2.0 and newer). The port required updates to the Python/C++ integration. Therefore, instead of using \textit{Cython}, which led to excessive code duplication, the integration has been updated to use \texttt{pybind11}.

\subsection{Extending operator coverage}

Although Mirage defines a broader set of operators at the graph level, the existing Triton code-generation path did not provide sufficient coverage for the workloads considered in this study. We extended the Triton backend to support additional operators required by these workloads, including SiLU, Sigmoid, Softmax, Sin, and Cos.

\subsection{Verification}

There are two complementary approaches for verifying candidate graphs: formal verification and probabilistic verification. In our Ascend implementation, we support both approaches, but use the formal verifier as the default verification mechanism in our experiments.

The formal verifier is implemented using the Rust \texttt{egg} library~\cite{egg}, which operates on user-defined mathematical rules for operators. A computation graph is translated into a mathematical expression and verified using operator-specific equivalence rules defined in the verifier. Consequently, extending the superoptimizer with a new operator also requires adding the corresponding mathematical rule to the formal verifier.

The formal approach is efficient and works well for the current Ascend port, but its coverage depends on the set of explicitly encoded equivalence rules. It can therefore fail to recognize an optimization whose correctness follows from a mathematical equivalence that is not represented by one of the available rules. Thus, in the current implementation, formal verification provides the correctness guarantee for the candidate graphs used during the Ascend search, within the scope of the encoded rules.

The probabilistic verifier provides a complementary approach that does not require explicit equivalence rules for each transformation. For each operator, it computes a fingerprint by performing a simplified execution of the operation on selected inputs. The fingerprints of the individual operators are combined to obtain a fingerprint for the complete computation graph, which can then be compared with the fingerprint of the original graph. Matching fingerprints provide probabilistic evidence that the two graphs implement the same computation. In particular, this approach can recognize mathematically equivalent graphs even when their structures differ, as it compares their computed fingerprints rather than relying on explicitly encoded transformation rules. Unlike formal verification, however, matching fingerprints does not constitute a mathematical proof of equivalence.

We do not currently use the probabilistic verifier as the primary verifier on Ascend because computing fingerprints for all operators in a candidate graph on the CPU introduces substantial verification overhead. Since superoptimization evaluates a large number of candidate graphs, repeatedly executing the fingerprint computations for every operator becomes impractical on the host CPU. A natural extension is therefore to generate the fingerprint computations as Triton kernels and execute them directly on the target accelerator. This would move the verification workload from the CPU to the Ascend device, substantially reducing the associated overhead and making probabilistic verification more practical for the superoptimization search.

\section{Experimental Methodology}

The experiments are conducted on an Ascend 910B3 NPU with 64\,GB of memory. Each kernel is profiled for 1000 iterations, and correctness is verified using \texttt{torch.allclose} with a tolerance of $10^{-2}$.

The evaluation covers matrix multiplication, unscaled softmax, RMSNorm+MatMul, gated MLP with SiLU activation, FNN, and attention. The workloads range from individual operators to multi-operator computations, allowing us to evaluate both shape-dependent mapping decisions and graph-level optimization. Simple workloads, such as Matrix multiplication and Softmax expose partitioning and reduction behavior, while fused workloads test cross-operator optimization. FNN and attention provide larger search spaces and more demanding backend requirements. It is worth noting that simple operator can be mapped to a single customized kernel, while a composite graph may contain multiple customized kernels and fused operations within them. The complete set of evaluated configurations is summarized in Table~\ref{tab:benchmark}.

For workloads with available reference implementations, we use handwritten Triton kernels from~\cite{TritonAscend}. For RMSNorm+MatMul and Gated MLP, we use kernels generated by DeepSeek~\cite{Guo2025}, as no corresponding handwritten Triton implementations were identified. This choice is motivated by the demonstrated effectiveness of DeepSeek models for GPU kernel generation and optimization in KernelBench~\cite{kernelbench}. These generated implementations are therefore treated as performance baselines rather than hardware-optimal references. The grid configurations of the reference and generated kernels are not constrained to be identical.

\begin{table}[ht]
    \centering
    \includegraphics[width=\linewidth]{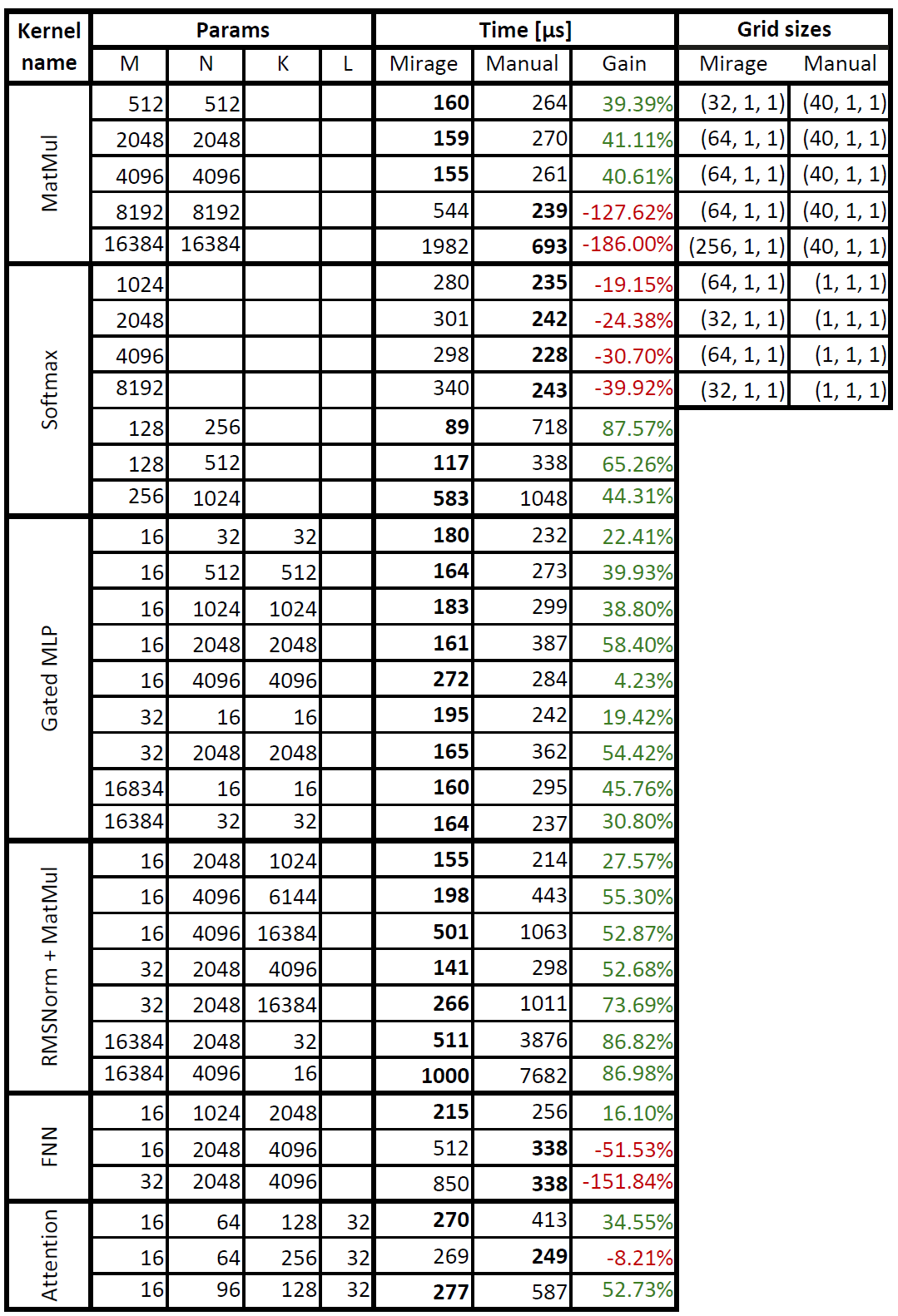}
    \caption{Performance comparison between superoptimizer-generated and manually optimized Triton kernels on Ascend 910B3.}
    \label{tab:benchmark}
\end{table}

\section{How well does graph superoptimization transfer?}

The results show that graph superoptimization transfers to Ascend, but that performance portability is strongly dependent on workload shape and structure. The generated kernels outperform the reference for several configurations, including both simple operators and multi-operator computations, while substantial regressions occur for other shapes. Thus, the graph representation and search infrastructure retain useful optimization capability on Ascend, but the current target mapping does not generalize uniformly across the workload space.

\subsection{Performance is highly shape-dependent}

Matrix multiplication provides the clearest example of shape-dependent performance (Table~\ref{tab:benchmark}). For the three smaller configurations, the generated kernels substantially outperform the handwritten Triton reference, achieving 39.4\%--41.1\% higher performance. This advantage reverses for larger matrices: at $N=8192$, the generated kernel is 127.6\% slower than the reference, and at $N=16384$, it is 186.0\% slower.

The computation remains a single MatMul across all configurations, so the change cannot be attributed to increasing graph complexity. Instead, it demonstrates that the current search identifies effective mappings for some tensor dimensions but not others. The selected grid configurations also change with matrix size, suggesting that the search makes substantially different scheduling decisions as the workload grows.

Softmax exhibits a similar shape dependence (Table~\ref{tab:benchmark}). For the one-dimensional configurations with $K=1024$--$8192$, the generated kernels are 19.2\%--39.9\% slower than the reference, whereas the remaining configurations show improvements of 44.3\%--87.6\%. The optimizer can therefore discover efficient reduction mappings, but their effectiveness depends strongly on tensor shape.

Together, these results show that performance portability cannot be characterized by the operator alone. Tensor dimensions influence the quality of the generated mapping even when the underlying computation is unchanged.

\subsection{Complex graphs remain optimizable}

Increasing graph complexity does not necessarily prevent successful optimization. FNN improves over the reference by 16.1\% for one configuration, but becomes 51.5\% and 151.8\% slower for two larger configurations. Attention shows a more balanced result, with two of three configurations outperforming the reference by 34.6\% and 52.7\%, while the remaining configuration is 8.2\% slower (Table~\ref{tab:benchmark}).

These results indicate that the graph representation remains expressive enough to optimize multi-stage computations on Ascend, but that the resulting performance becomes increasingly sensitive to target-specific lowering and scheduling decisions.

\subsection{Fused workloads demonstrate the potential of graph-level optimization}

\begin{figure*}[ht]
    \centering
    \begin{subfigure}[t]{\textwidth}
        \includegraphics[width=\linewidth]{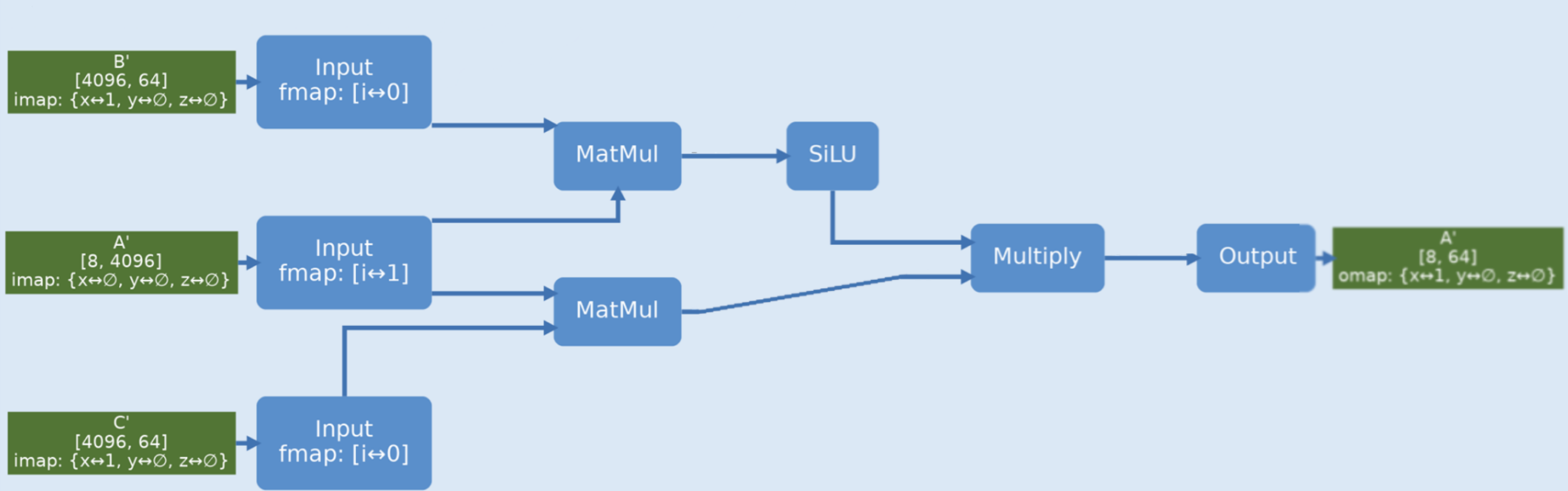}
        \caption{Before superoptimizer}
        \label{fig:MMbefore}
    \end{subfigure}
    \hfill
    \begin{subfigure}[t]{\textwidth}
        \includegraphics[width=\linewidth]{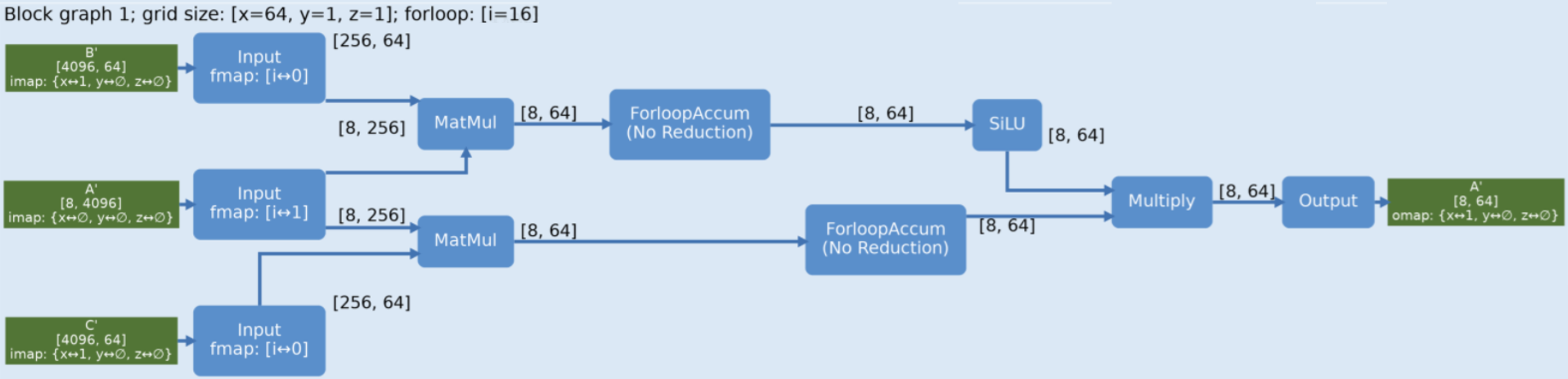}
        \caption{After superoptimizer}
        \label{fig:MMafter}
    \end{subfigure}
    \caption{Transformation of a Gated MLP computation graph before and after graph superoptimization.}
    \label{fig:beforevsafter}
\end{figure*}

Figure~\ref{fig:beforevsafter} illustrates this transformation for a Gated MLP workload. The optimized graph uses different tensor shapes, grid configuration, and accumulation structure while preserving the original computation.

RMSNorm+MatMul improves over its reference implementation in all seven tested configurations, with gains of 27.6\%--87.0\%, while Gated MLP improves in all nine configurations, with gains of 4.2\%--58.4\% (Table~\ref{tab:benchmark}).

These results demonstrate that the transferred graph-level optimization machinery can exploit cross-operator opportunities on Ascend. However, because these workloads use generated rather than handwritten reference implementations, they provide evidence of successful graph-level optimization rather than equivalent evidence of performance portability against expert-written kernels.

Overall, the results do not support a monotonic relationship between graph complexity and performance. Instead, successful transfer depends on the interaction between workload shape, graph structure, and the target-specific decisions required by the compilation pipeline.

\section{Why does performance portability break?}

The results show that the portability gap is not caused by a single component. Instead, several architectural assumptions remain embedded in the optimization and code-generation pipeline.

\subsection{Tensor partitioning remains CUDA-oriented}

Superoptimizer performs hierarchical search at the \texttt{LV\_KERNEL} and \texttt{LV\_THREADBLOCK} levels, exploring candidate operator compositions while partitioning and tiling tensors. The current partitioning formulation uses Z3~\cite{z3solver} to reason about tensor layouts under symbolic dimension constraints, but does not explicitly model Ascend-specific constraints such as Unified Buffer (UB) capacity.

This provides a likely explanation for the shape-dependent MatMul behavior observed above. The search can find effective partitions for some workloads, but the current formulation does not systematically optimize the partitioning space for the target memory hierarchy.

\subsection{Grid selection does not encode the AICore execution model}

Grid dimensions are part of the kernel configuration explored by the superoptimizer, but the search does not explicitly model their relationship to the available Ascend AICores. For example, on the target system, a grid size of 64 launches more programs than the 40 available AICores can execute concurrently, causing some programs to wait until AICores become available.

However, grid size alone does not explain the performance gap: matching the reference grid does not eliminate the regressions for the larger GEMMs. The more fundamental limitation is that the search lacks an explicit model of the target's execution resources and scheduling behavior.

\subsection{The Triton backend does not reproduce CUDA-level fusion transformations}

The CUDA and Triton backends lower the same optimized graph differently. In the CUDA backend, a dedicated \texttt{resolve\_tb\_fusion()} pass analyzes intermediate-tensor consumers and constructs eligible threadblock fusion chains. The Triton backend does not currently implement an equivalent transformation and instead directly traverses the operators in the threadblock graph during \texttt{transpile\_kn\_custom\_op()}.

As a result, the same superoptimizer representation can produce different kernel structures depending on the backend. The Triton backend generates one kernel per \texttt{KN\_CUSTOMIZED\_OP}, but does not perform the additional threadblock-level fusion introduced by \texttt{resolve\_tb\_fusion()} in the CUDA backend.

\subsection{Attention exposes search and lowering limitations}

Attention provides a boundary case where both search complexity and backend capabilities become important. Increasing the search depth can make the optimization prohibitively expensive, while representing additional operations such as scaling before softmax further enlarges the search space.

The lowering path introduces another limitation. The CUDA backend provides specialized handling for online softmax, which is not currently reproduced by the Triton backend. The Ascend implementation can therefore express and optimize several attention variants, but does not reproduce the complete FlashAttention-style transformation as reported in the original paper \cite{wu2025mirage}.

\subsection{Search and verification are not architecture-neutral in cost}

The search explores a combinatorial space of candidate graphs, with additional dimensions introduced by customized operators and tensor partitioning. Recent work has identified tensor partitioning as a major source of search-space explosion and proposed reformulating nested partitioning loops as explicit optimization variables~\cite{wu2026prism}. Verification adds further computational cost. The current Ascend implementation uses formal verification because probabilistic verification requires CPU-based fingerprint computation that is too slow for the intended workload.

The resulting portability issue is therefore broader than CUDA-specific assumptions: the practical cost of the complete search-and-verification pipeline depends on how these computations are executed on the target system.

\section{Lessons for portable accelerator optimization}

The Ascend adaptation demonstrates that a significant part of the graph superoptimization pipeline can be reused across different accelerator architectures. The graph representation, search framework, verification infrastructure, and Triton code generation enable the creation of functional kernels that, for several workloads, outperform available reference implementations.

At the same time, the results indicate that a reusable representation alone does not guarantee performance portability. To achieve consistent performance, decisions must take into account hardware-specific characteristics regarding tensor partitioning, execution resources, lowering, fusion, and search space management. Several design principles emerge from the adaptation.

\subsection{A portable IR does not imply a portable optimization space}

A common graph representation can preserve the structure of tensor computations across architectures without making the resulting optimization decisions portable. Effective partitions, tiles, schedules, and memory layouts depend on target-specific execution and memory resources. In the current Ascend adaptation, some of these mechanisms are only partially supported, limiting the ability of the optimizer to exploit the target architecture effectively.

Portable systems should therefore separate the reusable graph representation and search infrastructure from target-specific cost models and optimization mechanisms. In particular, target-specific partitioning and execution-resource modeling should be made explicit rather than implicitly inherited from the original CUDA-oriented design.

\subsection{Workload shape is part of the optimization problem}

The experiments show that optimization quality cannot be characterized independently of tensor dimensions. Different shapes expose different partitioning, tiling, reduction, and parallelization opportunities, so a configuration that is effective for one shape may be ineffective for another.

This observation also affects the evaluation of performance portability. Performance should therefore be evaluated over representative shape distributions rather than a single configuration. The optimizer should likewise account for workload shape when exploring the optimization space.

\subsection{Optimization must survive the entire compilation path}

A transformation discovered by the optimizer is useful only if subsequent lowering stages preserve and exploit it. The differences between the CUDA and Triton transpilers demonstrate that fusion and specialized transformations cannot be considered properties of the intermediate representation alone.

Furthermore, not every failed optimization can be attributed to the superoptimizer. Some generated Triton kernels are structurally correct but fail during later compilation, reflecting limitations or incompatibilities in the current compilation pipeline. At the same time, the experiments also point to internal limitations of the optimization pipeline, in particular the lack of explicit hardware modeling and differences between the CUDA and Triton lowering paths.

Search and lowering should therefore be designed as an end-to-end optimization pipeline, with target-specific stages made explicit and replaceable. This includes target-specific fusion, lowering, cost estimation, and verification mechanisms.

\subsection{Hardware-dependent components should be modular}

The results suggest that performance portability does not require making the entire optimizer independent of the hardware architecture. Instead, hardware-dependent components should be explicit and modular.

Such an architecture would allow target-specific partitioning, execution-resource modeling, cost estimation, fusion, and verification to be adapted to different accelerators while preserving the common graph representation and search engine. This separation would retain the benefits of a reusable superoptimization framework while allowing the optimization space and compilation path to reflect the characteristics of each target architecture.

\section{Conclusion}

This work demonstrates that graph-level superoptimization can be transferred from CUDA to the Huawei Ascend 910B3 through Triton. The adapted system generates competitive kernels and, for several workloads, substantially outperforms the available reference implementations.

However, the results also show that functional portability does not imply performance portability. Performance depends on workload shape and on target-specific decisions concerning tensor partitioning, execution resources, backend lowering, fusion, and search complexity.

The main lesson is that portable superoptimization should separate reusable graph-level infrastructure from accelerator-specific optimization mechanisms. Triton can provide an effective portability layer, but consistent performance still requires explicit target-aware modeling of the decisions that determine how the optimized graph is mapped to hardware.

Transfer succeeds when the target workload exposes optimization opportunities that the existing search and lowering pipeline can represent efficiently; it breaks when performance depends more strongly on target-specific partitioning, reduction handling, or backend-specific fusion.


\bibliographystyle{ACM-Reference-Format}
\bibliography{sample-base}

\end{document}